\documentclass[12pt]{article}
\pdfoutput=1
\usepackage{graphicx}
\usepackage{a4wide}
\usepackage{amssymb}
\begin{document}

\vskip .6cm

\centerline{{\LARGE \bf Baryonic symmetries in $AdS_4/CFT_3$: an overview}}

\medskip

\vspace*{4.0ex}

\centerline{
{\large \bf Diego Rodriguez-Gomez}\footnote{drodrigu@physics.technion.ac.il}}

\vspace*{4.0ex}

\begin{center}

\vspace*{4.0ex}
{\large Department of Physics\\
Technion, Haifa, 3200, Israel\\}
\vspace*{3.0ex}

{and}

\vspace*{3.0ex}
{\large Department of Mathematics and Physics\\
University of Haifa at Oranim, Tivon, 36006, Israel\\
}
\end{center}

\vspace*{5.0ex}

\centerline{\bf Abstract} \bigskip

Global symmetries play an important role in classifying the spectrum of a gauge theory. In the context of the $AdS/CFT$ duality, global baryon-like symmetries are specially interesting. In the gravity side, they correspond to vector fields in $AdS$ arising from KK reduction of the SUGRA $p$-form potentials. We concentrate on the $AdS_4/CFT_3$ case, which presents very interesting characteristic features. Following arXiv:1004.2045, we review aspects of such symmetries, clarifying along the way some arguments in that reference. As a byproduct, and in a slightly unrelated context, we also study $\mathcal{Z}$ minimization, focusing in the HVZ theory.

\newpage

\setcounter{footnote}{0}

\tableofcontents

\section{Introduction}

Over the last few years there has been considerable progress towards understanding the $AdS_4/CFT_3$ duality \cite{Maldacena:1997re}. The maximally supersymmetric example of this duality corresponds to the $AdS_4\times S^7$ space. This space arises as the near-horizon region of the background sourced by a stack of M2 branes moving in $\mathbb{C}^4$. Conversely, standard decoupling limit arguments show that a dual description is given by the $CFT_3$ on the worldvolume of the M2 branes. Following on the seminal work in \cite{Gustavsson:2007vu, Bagger:2007vi}, Aharony, Bergman, Jafferis and Maldacena (henceforth ABJM) constructed in \cite{Aharony:2008ug} what it is by now agreed to be the field theory dual to $N$ M2 branes probing the $\mathbb{C}^4/\mathbb{Z}_k$ singularity, of which the maximally SUSY example is the $k=1$ case. 

Since then, much activity has been devoted to further understand this duality in less supersymmetric cases. While there are purely theoretical reasons for that --as constructing and understanding dual pairs shedding information on both field theoretic and gravitational aspects--, a number of potential applications, in particular in what it has been dubbed the $AdS/CMT$ duality, have been recently considered. 

These less supersymmetric examples arise from M2 branes probing more involved singularities, which generically have a rich topological structure. In particular, supergravity $p$-form potentials can be KK reduced on these topologically non-trivial cycles giving rise to vector fields in $AdS$. In turn, these are related to global symmetries of the dual $CFT_3$. On general grounds, global symmetries play an important role in classifying the spectrum of a theory. Furthermore, they are also expected to be relevant from the point of view of potential applications of $AdS/CFT$. It is thus important to understand them in the context of the $AdS_4/CFT_3$ duality. 

Of particular interest are the global baryonic symmetries. These are abelian symmetries whose charged states have dimensions $\mathcal{O}(N)$. As such, they cannot correspond to KK states ($\Delta\sim\mathcal{O}(1)$), and must be dual to wrapped branes. Thus, they must be associated to the non-trivial topology of the cone where the M2 move. Indeed, as mentioned, non-trivial topology allows for the supergravity $p$-forms to wrap on cycles leading to gauge fields in $AdS_4$ as potential duals to these baryonic symmetries. However, as we will discuss below, following \cite{Benishti:2010jn} (see also \cite{Klebanov:2010tj}) the fate of these bulk fields, and their boundary duals, is remarkably different than the $AdS_5$ case (see \textit{e.g.} \cite{Martelli:2008cm} and references therein for an account of this case). In this short review we discuss several aspects of these symmetries by extracting as much information as possible from the gravity side of the correspondence. We start in section \ref{general} with a lightning overview of some relevant facts about the $AdS_4/CFT_3$ duality. We then turn in section \ref{chicha} to the baryonic symmetries of interest. In section \ref{example} we suggest an application to a particularly interesting geometry, in particular slightly clarifying arguments presented in \cite{Benishti:2010jn}. As a by-product, in the appendix we apply $\mathcal{Z}$-minimization to the HVZ theory.

\section{M2 branes probing $CY_4$: general aspects}\label{general}

As discussed in the introduction, the $AdS_4/CFT_3$ duality arises as the near horizon limit of a stack of M2 branes probing a conical geometry. In fact, the low energy limit of the M2 brane worldvolume theory must supply the CFT side of the correspondence, according to the usual decoupling limit arguments \cite{Maldacena:1997re}. 

The best understood case is that of M2 branes in flat space, when the near horizon region is the maximally supersymmetric $AdS_4\times S^7$ space. In turn, the dual field theory is the $U(N)\times U(N)$ Chern-Simons theory with levels $(1,\,-1)$ and particular matter content constructed in \cite{Aharony:2008ug}. This theory arises as a member of a whole family of $\mathcal{N}=6$ SCFT's with levels $(k,\,-k)$ \cite{Aharony:2008ug, Benna:2008zy}. For generic $k$ the moduli space is the orbifold  $\mathbb{C}^4/\mathbb{Z}_k|_{(1,\,1,\,-1,\,-1)}$.  It is only for $k=1,\,2$ that there is a quantum-mechanical enhancement to $\mathcal{N}=8$ due to special properties of monopole operators. Conversely, the gravity side of the duality is provided by the near horizon region of the background sourced by a stack of M2 branes proving this orbifold, namely $AdS_4\times S^7/\mathbb{Z}_k$. The $\mathbb{Z}_k$ orbifold acts by quotienting the $U(1)$ fiber of the fibration $S^7\sim S^1\hookrightarrow \mathbb{P}^3$. In fact, in the large $k$ limit, the fiber shrinks and the geometry is better understood as the IIA $AdS_4\times \mathbb{P}^3$ background with suitable fluxes to preserve 24 supersymmetries. From this perspective, the vector of CS levels in gauge group space specifies the $U(1)$ dual to the M-theory circle. Indeed, diagonal monopole operators, charged under this $U(1)$, become the KK states of the reduction, \textit{i.e.} the D0 branes \cite{Aharony:2008ug}.

It is crearly greatly desirable to understand the $AdS_4/CFT_3$ duality in the generic case, where the M2 branes probe less symmetric spaces $X$. On general grounds, the radius/energy relation of $AdS/CFT$ requires the manifold $X$ to be a cone over a 7-dimensional base $Y$, \textit{i.e.} $ds^2(X)=dr^2+r^2\,ds(Y)^2$. Then the appropriate 11-dimensional supergravity solution corresponding to $N$ M2 branes located at the tip of $X$ is

\begin{equation}
\label{11dmetric}
ds_{11}^2=h^{-2/3}\, ds^2(\mathbb{R}^{1,\,2})+h^{1/3}\,ds^2(X)\, ,\qquad G=d^3x\wedge dh^{-1}\, ,\qquad h=1+\frac{R^6}{r^6}\, .
\end{equation}
In the near horizon limit, and upon defining $z=R^2/r^2$, the space becomes a Freund-Rubin product space between $AdS_4$, whose metric in Poincare coordinates is

\begin{equation}
\label{z_coordinate}
ds^2(AdS_4)^=\frac{dz^2+dx(\mathbb{R}^{1,2})^2}{z^2}\,;
\end{equation}
and the base $Y$ 

\begin{equation}
ds_{11}^2=R^2\Big(\frac{1}{4}\,ds^2(AdS_4)+ds^2(Y)\Big)\, ,\qquad G=\frac{3}{8}\, R^3\,{\rm Vol}(AdS_4)\ .
\end{equation}

Furthermore, the flux quantization condition leads to the relation 

\begin{equation}
R=2\,\pi\,\ell_P\,\Big(\frac{N}{6\,{\rm Vol}(Y)}\Big)^{\frac{1}{6}}\, .
\end{equation}

On the other hand, constructing the corresponding dual field theories has proved remarkably difficult. Only in the last few years we have seen big progress along these lines. From the CFT point of view, general field theory arguments, discussed for the 3d case at hand in  \cite{Berenstein:2009sa}, show that theories with $\mathcal{N}\geq 2$ are of special interest due to the existence of a $U(1)_R$ symmetry. This symmetry endows the moduli space of a graded structure which allows to classify chiral operators according to their R-charge; which equals, in virtue of the superconformal algebra, their scaling dimension. At the same time, it automatically implies that the moduli space has a cone-like structure. We will thus demand $\mathcal{N}\geq 2$, which in turn requires, on general grounds  \cite{Acharya:1998db}, the M2 branes to move in spaces of at most $SU(4)$ holonomy. Following the ABJM example, it is natural to consider Chern-Simons-matter theories as potential SCFT duals. As shown in  \cite{Gaiotto:2007qi}, $\mathcal{N}\geq 3$ fixes the superpotential couplings to be proportional to the CS levels, thus almost ensuring conformal invariance. However, for our purposes we will be mostly interested in the less restrictive but yet tractable (due to the existence of $U(1)_R$) $\mathcal{N}=2$ case, where the dual geometry is strictly $CY_4$ (that is, $Y$ is Sasaki-Einstein), which we will further assume toric. While we refer the reader to the standard literature for a thorough introduction to toric geometry (for a physics related discussion, see \textit{e.g.} \cite{Martelli:2004wu}), let us briefly highlight, for completeness, the basic ideas. The cone ${\cal C}(Y)$ is \emph{toric} if it can be seen as a $U(1)^4$ fibration over a  polyhedral cone in $\mathbb{R}^4$ . This polyhedral cone defined as the convex set of the form $\bigcap\{\mathbf{x}\cdot \mathbf{v}_\alpha\geq 0\}\subset\mathbb{R}^4$, where $\mathbf{v}_\alpha\in \mathbb{Z}^4$ are integer vectors. The Calabi-Yau condition implies that, with a suitable choice of basis, we can write $\mathbf{v}_\alpha=(1,\mathbf{w}_\alpha)$, with $\mathbf{w}_\alpha\in\mathbb{Z}^3$. If we plot these latter points in $\mathbb{R}^3$ and take their convex hull, we obtain the \emph{toric diagram}. In fact, the toric diagram contains all the relevant information about the $CY_4$ geometry. 

\begin{figure}[ht]
\begin{center}
\includegraphics[scale=.40,angle=-90]{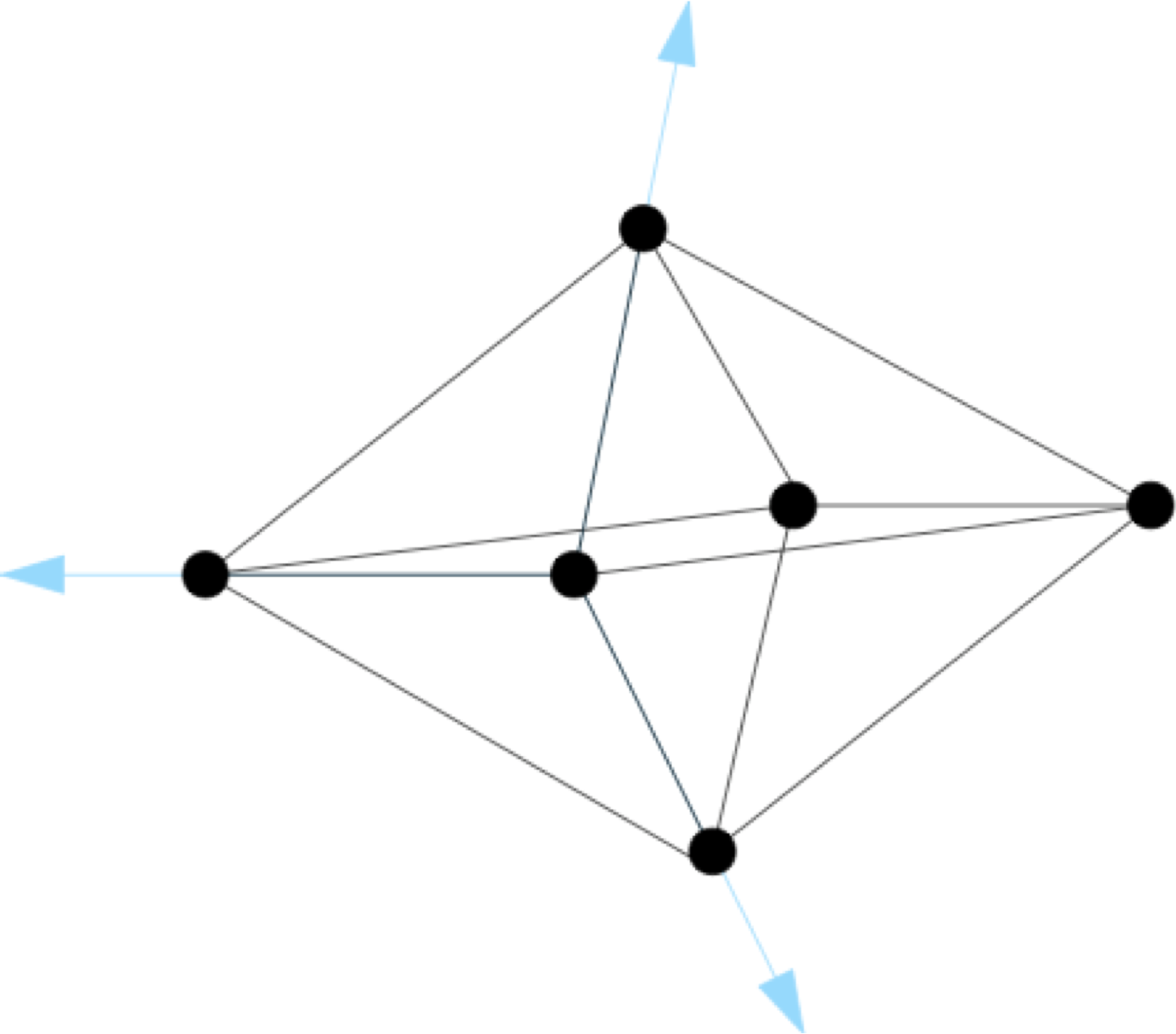}
\end{center}
\caption{The toric diagram for $\mathcal{C}(Q^{111})$.}
\label{fig:toricdiagramQ111}
\end{figure}

As shown in \cite{Jafferis:2008qz, Martelli:2008si, Hanany:2008cd} and briefly reviewed in section \ref{example}, toric manifolds naturally arise as moduli space of $\mathcal{N}=2$ CS-matter quiver gauge theories with toric superpotentials\footnote{By toric $W$ we mean a $W$ where each field appears exactly twice, one time in a monomial with $+$ sign, another time in a monomial with sign $-$.} whose levels add up to zero. Furthermore, very much like in ABJM, the CS level vector in gauge space selects the M-theory circle, which at generic level is quotiented. Thus, the actual moduli space of these $\mathcal{N}=2$ Chern-Simons-matter theories is a certain $\mathbb{Z}_k$ quotient of the toric $CY_4$. In section \ref{example} we will study in more detail one such example, conjectured to be dual to the cone over $Q^{111}$, whose toric diagram we show in fig. (\ref{fig:toricdiagramQ111}).

We should note that, as opposed to the ABJM case, in the $\mathcal{N}=2$ cases this circle generically collapses as one moves on the base of the cone. This motivates the recently appeared proposals \cite{Jafferis:2009th, Benini:2009qs} involving fundamental matter as well as bifundamental fields, as, on general grounds, associated to these collapsing locii there can be extra flavor branes in the IIA reduction. 

Yet one more warning note is in order. While the construction \cite{Jafferis:2008qz, Martelli:2008si, Hanany:2008cd} yields to toric $CY_4$ classical abelian moduli spaces, it yet remains to be understood wether at the non-abelian quantum level these theories are indeed SCFT's.  Only very rencently a manageable criterion to determine wether a 3d theory flows to an IR fixed point, which amounts to the minimization of the partition function $\mathcal{Z}$, has been proposed in \cite{Jafferis:2010un} (see also \cite{Hama:2010av}). One particular example where to put this at practice is the HVZ theory \cite{Hanany:2008fj}.  While at the classical abelian level the moduli space is $\mathbb{C}^2/\mathbb{Z}_k\times \mathbb{C}^2$, a more careful analysis  \cite{RodriguezGomez:2009ae}  shows that the chiral ring (studied at large $k$ to avoid subtleties with monopole operators) contains completely unexpected non-abelian branches while there is no trace of the neccesary $SO(4)_R$ symmetry of the generically $\mathcal{N}=4$ orbifold. In fact, as shown in \cite{Choi:2008za}, the superconformal index fails to meet the gravity expectations. Indeed, as briefly discussed in the appendix, when the $\mathcal{Z}$-minimization is applied to the HVZ theory it suggests that for no $k$ it can be dual to the ABJM model. In \cite{Kim:2010vwa} a variant of the theory with explicit $\mathcal{N}=3$ SUSY and no extra branches in the chiral ring was considered, finding however that the index computation was still in disagreement with the expectations.

\section{Global symmetries in $AdS_4/CFT_3$ and their spontaneous breaking}\label{chicha}

We have so far discussed generic aspects of the $AdS_4/CFT_3$ duality. As described, the cases of interest are those where a stack of M2 branes probes a $CY_4$ cone. In turn, these cones generically have a non-trivial topology, in particular containing $b_2(Y)\ne 0$ 2-cycles. This allows the fluctuations of the supergravity potentials to wrap on them yielding to vector fields on $AdS_4$. In fact, due to Poincare duality ${\rm dim}\,H_5(Y)={\rm dim}\,H_2(Y)=b_2(Y)$. We can then introduce a set of dual harmonic five-forms $\alpha_1,\cdots \alpha_{b_2(Y)}$ and consider 6-form potential fluctuations of the form 

\begin{equation}
\delta C_6=\frac{2\pi}{T_5}\, \sum_{I=1}^{b_2(Y)}\,\mathcal{A}_I\wedge \alpha_I\, .
\end{equation}
Upon KK reduction, this gives rise to $b_2(Y)$ massless gauge fields $\mathcal{A}_I$ in $AdS_4$. These fields sit in certain multiplets, known from the supergravity point of view as \emph{Betti multiplets} (see \textit{e.g.}   \cite{D'Auria:1984vv}).

In the context of the $AdS_5/CFT_4$, these Betti symmetries correspond to global baryonic symmetries on the field theory side. In fact, these arise from the $U(1)$ factors inside the $\prod U(N)$ total gauge group, which in 4d are IR free. It is possible to show that indeed the $b_2$ non-anomalous such $U(1)$'s --which appear as global baryonic symmetries-- are identified with these Betti multiplets (see \textit{e.g.} \cite{Martelli:2008cm} and references therein for a comprehensive discussion). 

In turn, in the $AdS_4/CFT_3$ case the role of this symmetries must be different. This can be inferred from general field theory arguments, as they clearly cannot arise from decoupled $U(1)$ factors, which are not IR free in 3-dimensions. Nevertheless, due to their origin, similar to the $AdS_5$ case, we will still refer to them as baryonic symmetries.\footnote{When referring to the ABJM theory the difference $U(1)$ gauge field is sometimes also called baryonic $U(1)$, mirroring the Klebanov-Witten terminology --recall that ABJM is described by the same quiver an superpotential as the Klebanov-Witten theory, only in one dimension less and adding CS for the gauge groups--. We stress that our baryonic symmetries are very different from this one, which is basically the M-theory circle.} Since on general grounds global symmetries are of much help in classifying the spectrum of a gauge theory, the study of such baryonic-like $U(1)$'s is indeed of much interest. Let us now turn to the supergravity side to extract as much information as possible about these symmetries and their implication in the dual field theory.

Let us note that while the $CY_4$ might have other types of cycles, only 2-cycles (and the Poincare-dual 5-cycles) are relevant for our discussion. As discussed in  \cite{Benishti:2010jn}, the toric $CY_4$ of interest can typically have additional 6-cycles, which manifest themselves as internal points in the toric diagram. Nevertheless, it is clear that these will not lead to vector fields in $AdS_4$ upon KK reduction of SUGRA $p$-forms on them, and so their role must be different than that of 2- and 5-cycles. In fact, as briefly discussed in  \cite{Benishti:2010jn}, it appears that these 6-cycles can yield to non-perturbative corrections to superpotentials, as euclidean 5-branes can be wrapped on them. Since we will be mostly concerned with global baryon-like symmetries, we will not touch upon these 6-cycles and focus for the rest of the contribution on 2- and 5-cycles.

Finally, making use of results in  \cite{lerman}, in \cite{Benishti:2010jn} it was argued that the number of such two-cyles is given by $b_2(Y)=d-4$, being $d$ the number of external points in the toric diagram. While this result is strictly valid only for isolated singularities, we note it coincides with the conjecture in \cite{Davey:2009sr,Davey:2009qx}. We note that, as discussed above, internal points, being related to 6-cycles over which no SUGRA $p$-form yields to an $AdS_4$ vector upon KK reduction, are not related to baryonic symmetries. Conversely, the $d-4$ number of such symmetries does not depend on the number of internal points.

\subsection{Gauge fields in $AdS_4$}

The $b_2(Y)$ vector fields satisfy, at the linearized level, Maxwell equations in $AdS_4$.\footnote{The vector fields arising from KK reduction correspond to abelian bulk gauge fields, and thus will correspond to global/gauged $U(1)$ boundary symmetries. In fact, as discussed in the main text, wrapped branes behave as sources of this abelian theory. Thus, we do not expect any non-abelian enhancement. Note that this argument is strictly applicable to isolated singularities.}    Furthermore, these $b_2(Y)$ copies of 4d E\&M generically contain both electric and magnetic pointlike sources in $AdS_4$. From the 11-dimensional point of view, these pointlike electrons and monopoles will become wrapped branes, and their role will be crucial in the following. 

Let us analyze more in detail E\&M in $AdS_4$. In fact, we will keep the discussion generic, and consider a vector field in $AdS_{d+1}$. We can set $A_z=0$ away from the sources. Then, using the straightforward generalization to $AdS_{d+1}$ of the coordinates in (\ref{z_coordinate}), the bulk equations of motion set
\begin{equation}
\label{vectAdS4behavior}
\label{gauge_field_In_AdS4}
A_{\mu}=a_{\mu}+j_{\mu}\, z^{d-2} \ ,
\end{equation}
where the $a_{\mu},\, j_{\mu}$ satisfy the free Maxwell equation in the boundary directions. Furthermore, Lorentz gauge for these is automatically imposed. In fact, this can be naturally interpreted as fixing bulk Coulomb gauge upon regarding $z$ as the time coordinate. The condition $A_z=0$ away from the source is then the standard radiation gauge in that context.  

The $AdS/CFT$ duality requires specifying the boundary conditions for the fluctuating fields in $AdS$. In particular, and crucially different to $AdS_5$, vector fields in $AdS_4$ admit different sets of boundary conditions \cite{Witten:2003ya,Leigh:2003ez, Marolf:2006nd} leading to different boundary CFT's. Coming back to (\ref{vectAdS4behavior}), it turns out that in $d<4$ both behaviors have finite action, and thus can be used to define a consistent $AdS/CFT$ duality. Furthermore, the fluctuations $a_{\mu},\, j_{\mu}$ are naturally identified, according to the $AdS/CFT$ rules,  with a dynamical gauge field and a global current in the boundary respectively. In accordance with this identification, eq. (\ref{gauge_field_In_AdS4}) and the usual $AdS/CFT$ prescription shows each field to have the correct scaling dimension for this interpretation: for a gauge field $\Delta(a_{\mu})=1$, while for a global current $\Delta(j_{\mu})=2$. 

Let us now concentrate on the case of interest $d=3$, where both quantizations are allowed. In order to have a well-defined variational problem for the gauge field in $AdS_4$ we should be careful with the boundary terms when varying the action. In general, we have
\begin{equation}
\delta S = \int \Big\{ \frac{\partial \sqrt{\det g}\,\mathcal{L}}{\partial A_M}-\partial_N\frac{\partial \sqrt{\det g}\,\mathcal{L}}{\partial\partial_NA_M}\Big\}\, \delta A_M +\partial_N\Big\{ \frac{\partial\sqrt{\det g}\, \mathcal{L}}{\partial\partial_NA_M}\, \delta A_M\Big\} \ .
\end{equation}
The bulk term gives the equations of motion whose solution behaves as (\ref{gauge_field_In_AdS4}). In turn, the boundary term can be seen to reduce to 
\begin{equation}
\delta S_B=- \frac{1}{2}\,\int_{\mathrm{Boundary}} \, j_{\mu}\,\delta a^{\mu} \, .
\end{equation}
Therefore, in order to have a well-posed variational problem, we need to demand $\delta a_{\mu}=0$; that is, we need to impose boundary conditions where $a_{\mu}$ is fixed in the boundary. 

On the other hand, since in $d=3$ both behaviours for the gauge field have finite action, we can consider adding suitable boundary terms such that the action becomes \cite{Marolf:2006nd}
\begin{equation}
\label{dynamical_a}
S=\frac{1}{4}\int \sqrt{\det g}\, F_{AB}\, F^{AB} +\frac{1}{2}\,\int_{\mathrm{Boundary}}\, \sqrt{\det g}\, A^{\mu} \, F_{z\mu}|_{\mathrm{Boundary}}  .
\end{equation}
The boundary term is now 
\begin{equation}
\delta S_B=\frac{1}{2}\,\int_{\mathrm{Boundary}} \,  a_{\mu}\delta j^{\mu} \, ,
\end{equation}
so that we need to impose the boundary condition $\delta j_{\mu}=0$; that is, fix the boundary value of $j_{\mu}$. 

The radiation-like gauge $A_z=0$ suggests to interpret $z$ as the time direction. Defining then the usual electric and magnetic fields $\vec{B}=\frac{1}{2}\epsilon^{\mu\nu\rho}\, F_{\nu\rho}$ and $\vec{E}=F_{\mu z}$, we have
\begin{equation}
B^{\mu}=\epsilon^{\mu\nu\rho}\partial_{\nu}a_{\rho}+\epsilon^{\mu\nu\rho}\partial_{\nu}j_{\rho}\, z\,, \qquad E^{\mu}=j^{\mu}\, z^2 \ .
\end{equation}
In terms of these, the two sets of boundary conditions correspond, on the boundary, to either setting $E_{\mu}=0$ while leaving $a_{\mu}$ unrestricted, or setting $B_{\mu}=0$ while leaving $j_{\mu}$ unrestricted. To be more explicit, recalling the $AdS/CFT$ interpretation of $a_{\mu},\,j_{\mu}$, the quantization $E_{\mu}=0$ is dual to a boundary CFT where the $U(1)$ gauge field is \emph{dynamical}; while the quantization $B_{\mu}=0$ is dual to a boundary CFT where the $U(1)$ is ungauged and is instead a \emph{global symmetry}. Furthermore, as discussed in \cite{Klebanov:1999tb} for the scalar counterpart, once the improved action is taken into account the two quantizations are Legendre transformations of one another \cite{Klebanov:2010tj}, as can be seen by \textit{e.g.} computing the free energy in each case. 

In turn, this has an important consequence for the spectrum of electrons and monopoles in this 4d E\&M --which of course come wrapped branes from an 11-dimensional point of view--. Let us consider an M5 brane wrapped in one of the $b_2(Y)$ 5-manifolds $\Sigma_5\subset Y$. From the $AdS_4$ point of view, this brane looks like a pointlike electric charge for the corresponding vector field. On the other hand, the linearized $C_6$ fluctuation which such brane sources must be of the form $\delta C_6\sim f(z)\, dt\wedge {\rm Vol(\Sigma_5)}$. Upon reduction this precisely yields  to $E_0\ne 0$ while $B_{\mu}=0$. Thus, it follows that wrapped M5 branes are only allowed upon choosing the quantization condition which fixes $a_{\mu}$. Conversely, dual wrapped M2 branes, though non-SUSY, would only be allowed upon choosing the boundary conditions which fix $j_{\mu}$. In turn, these boundary conditions do forbid the wrapped M5.

One can consider electric-magnetic duality in the bulk theory, which exchanges $E_{\mu}\leftrightarrow B_{\mu}$ thus exchanging the two boundary conditions for the $AdS_4$ gauge field quantization. This action translates in the boundary theory into the so-called $\mathcal{S}$ \textit{operation} \cite{Witten:2003ya}. This is an operation on three-dimensional CFTs with a global $U(1)$ symmetry, taking one such CFT to another. In addition, it is possible to construct a $\mathcal{T}$ \textit{operation}, which amounts, from the bulk perspective, to a shift of the bulk $\theta$-angle by $2\pi$. In fact, these two operations generate an $SL(2,\,\mathbb{Z})$ algebra transforming among the possible generalized boundary conditions (\cite{Witten:2003ya,Leigh:2003ez}).

\subsubsection{Wrapped branes in $AdS_4$ and baryonic operators}

 As the gauge symmetries in $AdS_4$ of interest arise from reduction of the SUGRA potentials, it is clear that no usual KK-state will be charged under them --the converse holds for the dual operators in the CFT side--. In turn, as described above, the relevant objects charged under them are M5 branes, which act as electric sources once the appropriate boundary conditions have been selected. Let us discuss these branes in more detail for the toric $CY_4$'s at hand. In these cases, an M5 brane wrapped on a five-manifold $\Sigma_5\subset Y$, such that the cone ${\cal C}(\Sigma_5)$ is a complex divisor in the K\"ahler cone ${\cal C}(Y)$, is supersymmetric and leads to a BPS particle propagating in $AdS_4$. As we argued in the previous subsection, since the M5 brane is a source for $C_6$, this particle is electrically charged under the $b_2(Y)$ massless $U(1)$ gauge fields $\mathcal{A}_I$. One might also consider M2 branes wrapped on two-cycles in $Y$. However, such wrapped M2 branes are not supersymmetric, as there are no calibrating 3-forms for the cone over the $\Sigma_2$ submanifold which they would wrap.

For toric manifolds there is a canonical set of wrapped M5 brane states, where ${\cal C}(\Sigma_5)$ are taken to be the toric divisors. In fact, the set of vectors defining the toric diagram introduced above is precisely the set of charge vectors specifying the $U(1)$ subgroups of $U(1)^4$ that have complex codimension one fixed point sets, and thus must correspond to the 5-manifolds where to wrap the M5 branes. To make this precise, in the $Q^{111}$ example the toric divisors correspond to the 6 external points in the toric diagram in fig.(\ref{fig:toricdiagramQ111}).

The standard rules of the $AdS/CFT$ prescription allow to identify these wrapped M5 branes, whenever the boundary conditions allow for them, with chiral operators in the dual field theory. In fact, as they correspond to non-perturbative states in supergravity, we should expect their scaling dimension to be of order $N$. In order to check this, we can consider changing to global coordinates for $AdS$, such that the energy of a particle in $AdS$ in units of $1/R$ is directly the scaling dimension in the field theory. For the wrapped branes under consideration it is straightforward to show that the action reduces to 

\begin{equation}
S=T_5\,{\rm Vol}(\Sigma_5)\, R^5\,\int dt\, \sqrt{\hat{g}}\,\hat{g}_{tt}\, ,
\end{equation}
where $\hat{g}$ stands for the $AdS_4$ metric in global coordinates. Thus, this is indeed describes  a mass $m=T_5\,R^5\,{\rm Vol}(\Sigma_5)$ particle in global $AdS_4$. Thus, through $AdS/CFT$, the dimension of the dual operator is

\begin{equation}
\label{Delta}
\Delta(\Sigma_5)=m\,R=T_5\,R^6\,{\rm Vol}(\Sigma_5)=N\,\frac{\pi}{6}\,\frac{{\rm Vol}(\Sigma_5)}{{\rm Vol}(Y)}\, .
\end{equation}
As the ratio of the volume of the 5-manifold to the ratio of $Y$ is an $\mathcal{O}(1)$ number, it follows that in fact these wrapped M5 branes must correspond to $\mathcal{O}(N)$ operators.

\subsection{Field theory perspective of Betti symmetries}

In the previous sections we have seen that the KK reduction of supergravity potentials must lead, on the boundary, to either a gauge or a global symmetry; depending on the choice of boundary conditions. This arises as, crucially, both boundary behaviors for gauge fields in $AdS_4$ are allowed; and it is the choice of boundary conditions what selects wether these bulk gauge fields correspond to a boundary gauge or global symmetry. Consistently, the choice of boundary conditions also determines which wrapped objects are allowed. Through $AdS/CFT$, as discussed in the previous subsection, these objects correspond to operators of dimension $\mathcal{O}(N)$.

On general grounds, the suitable CFT's dual to the toric geometries of interest will be $\prod U(N)$ gauge theories. These theories will contain a chiral ring consisting on a set of chiral operators with protected dimensions such that in the large $N$ limit they remain $\mathcal{O}(1)$. As their dimensions remain small, these operators must correspond to KK states in the gravity side. On the other hand, if a global baryonic symmetry is present in the theory, we expect baryon-like operators with dimensions $\mathcal{O}(N)$. The natural form of these operators is $\mathcal{B}={\rm det}X$, being $X$ a certain field charged under the corresponding baryonic symmetry. Conversely, these $\mathcal{O}(N)$ dimension operators must correspond to wrapped branes in the gravity dual, that is, the M5 branes wrapped on toric divisors we have just discussed. In turn, from the gravity analysis above, we learn that these branes are allowed once the suitable boundary conditions have been chosen --namely those fixing $a_{\mu}$ on the boundary and leaving a dynamical $j_{\mu}$, which has the correct properties for a global symmetry current--. On the other hand, the set of boundary conditions which do not allow for the wrapped M5 branes must correspond to a theory where the baryonic symmetry is gauged (instead of global). Consistently, the boundary $a_{\mu}$ is dynamical, which in fact has the correct features to be identified with a gauge field. In turn, being the $U(1)_B$ a gauged symmetry, the baryon-like operators would be forbidden because of gauge non-invariance; thus reflecting the lack of wrapped M5's. Therefore, for each baryonic symmetry we should expect two \emph{different} dual CFT's, each associated to a choice of boundary conditions, where the baryonic $U(1)$ symmetries are either gauged or global. We stress that these theories are different CFT's, related though by the gauging/ungauging of the $U(1)_B$'s. In fact, the gravity dual allows us to be more precise. As reviewed above, the exchange of the boundary conditions stands for the electric-magnetic duality of the $AdS_4$ E\&M. It is possible to enhance this action with yet another transformation so that we have an $SL(2,\,\mathbb{Z})$ action. Following \cite{Witten:2003ya} (see also \cite{Leigh:2003ez}), these bulk actions  translate in a precise way to the boundary CFT. Starting with a three-dimensional CFT with a global $U(1)$ current $j^{\mu}$, one can couple this global current to a background gauge field $A$ resulting in the action $S[A]$. The $\mathcal{S}$ operation then adds a BF coupling of $A$ to a new background field $B$ and at the same time promotes $A$ to a dynamical gauge field by introducing the functional integral over it; while the $\mathcal{T}$ operation instead adds a CS term for the background gauge field $A$:
\begin{equation}
\label{SL2Zaction}
\mathcal{S}:\, S[A]\,\rightarrow\, S[A]+\frac{1}{2\pi} \int B\wedge d A~,\qquad \mathcal{T}:\,S[A]\,\rightarrow\, S[A]+\frac{1}{4\pi}\int A\wedge d A \ .
\end{equation}
As shown in \cite{Witten:2003ya}, these two operations generate the group $SL(2,\mathbb{Z})$.\footnote{Even though we are explicitly discussing the effect of $SL(2,\mathbb{Z})$ on the vector fields, since these are part of a whole Betti multiplet we expect a similar action on the other fields of the multiplet. We leave this investigation for future work.} In turn, as discussed above, the $\mathcal{S}$ and $\mathcal{T}$ operations have the bulk interpretation of exchanging $E_{\mu}\leftrightarrow B_{\mu}$ and shifting the bulk $\theta$-angle by $2\pi$, respectively. It is important to stress that these actions on the bulk theory change the boundary conditions. Because of this, the dual CFTs living on the boundary are different.

\subsection{Spontaneous symmetry breaking}

We have seen that the choice of boundary conditions where we fix the boundary value of the bulk vectors arising from KK reduction of the supergravity potentials lead, on the CFT side, to global symmetries. On general grounds, we might then consider their spontaneous breaking to further test the consistency of the picture. In turn, generically, we should expect spontaneous symmetry breaking to correspond, in the gravity side, to Calabi-Yau resolutions of the cone \cite{Klebanov:1999tb} where an $S^2$ --of radius $b$-- is blown-up.

Upon resolution, the $CY_4$ will only be assymptotically conical. In fact, the first correction to the assymptotic cone-like metric generically goes like $r^{-2}$, which leads to the following behavior for the warp factor

\begin{equation}
h\sim \frac{R^6}{r^6}\,(1+\frac{b}{r^2}+\cdots)\, .
\end{equation}
Recalling the relation between the cone radial coordinate and the appropriate $AdS_4$ radial coordinate, according to the standard $AdS/CFT$ rules the subleading correction $\mathcal{O}(z^{-1})$ must be dual to a dimension 1 operator which acquires a VEV proportional to $b$. In fact, the natural candidate is the scalar component $\mathcal{U}$ in the global current multiplet, whose dimension is protected by supersymmetry to be 1. This operator is roughly the moment map of the $U(1)_B$ action, and is of the form

\begin{equation}
\mathcal{U}=\frac{1}{N}\sum_{{\rm charged\,fields}} {\rm Tr}\,q_{X_i}\,X_i\,X_i^{\dagger}\, .
\end{equation}
It is then clear that spontaneous symmetry breaking, triggered by a VEV of a scalar with charge $q_{X_i}$ under the $U(1)_B$, will give a VEV to $\mathcal{U}$. Furthermore, this VEV must trigger an RG-flow to a different fixed point. In turn, in the gravity side, much like in \cite{Klebanov:2007us}, upon using the appropriate radial coordinate, close to the branes the space develops an $AdS_4$ throat which stands for the IR fixed point.

\subsubsection{The order parameter for SSB}

The baryonic $U(1)_B$ symmetry is broken whenever a field $X$ charged under it takes a VEV. In particular, the $\mathcal{U}$ operator discussed above signals such breaking. However, a natural operator to consider is the associated baryon $\mathcal{B}={\rm det}X$, which, as discussed above, corresponds to a BPS particle in $AdS_4$ arising from a wrapped M5 brane on $\Sigma_5$. From the gravity perspective we can compute its VEV by considering the action $S_E$ of an euclidean brane which wraps the cone over $\Sigma_5$ --the so-called baryonic condensate--. Indeed, the $AdS/CFT$ dictionary allows to identify

\begin{equation}
\langle \mathcal{B}\rangle = e^{-S_E}\, .
\end{equation}
Let us concentrate on the modulus of the VEV, which comes from the exponential of the DBI action of the euclidean brane. Quite remarkably, as shown in \cite{Benishti:2010jn}, this contribution, which amounts to the warped volume of the cone over $\Sigma_5$, can be computed generically for the toric $CY_4$ of interest. Such warped volume is divergent, and it is then necessary to regulate it cutting off the integral at some large $r_c$. We refer to \cite{Benishti:2010jn} for the details of the computation. For the time being, let us quote the most relevant aspect of the result, namely that the modulus of the VEV is proportional to

\begin{equation}
\langle \mathcal{B}\rangle \sim z^{-\Delta(\Sigma_5)}\, .
\end{equation}
This result from supergravity can be seen as a prediction for the field theory dual. Indeed, if the expected dual operator is $\langle {\rm det} X\rangle$, we would expect its scaling dimension to be $N\,\Delta(X)$, so that $\Delta(X)=N^{-1}\,\Delta(\Sigma_5)$, in agreement with (\ref{Delta}).

\subsubsection{The emergence of the Goldstone particle and the global string}

In the preceding section we concentrated on the modulus of the VEV of the baryonic operator obtaining non-trivial expectations for the dual field theory. However, a complete picture of spontaneous symmetry breaking must involve the identification of the associated Goldstone boson. On general grounds, field theoretic spontaneous symmetry breaking can lead to cosmic strings around which such Goldstone boson would have a non-trivial monodromy. In fact, following the $AdS_5$ example \cite{Klebanov:2007cx}, in the gravity dual these strings can be easily identified as M2 branes wrapping the blown-up 2-cycle. Remarkably, these branes remain of finite tension at the bottom of the cone in the warped geometry (\ref{11dmetric}) where $ds^2(X)$ is replaced by the resolved cone metric. 

The finite tension M2 branes wrapped on the blown up cycle appear as a pointlike object in the Minkowski directions. In fact, in 3-dimensions they correspond to cosmic ``strings". In order to complete this picture, we must find the Goldstone boson winding around them. To that matter, we consider a 3-form linearized fluctuation \cite{Benishti:2010jn}

\begin{equation}
\delta\,C_3=A\wedge \beta\, ,
\end{equation}
where $\beta$ is a 2-form which, in the bottom of the cone, becomes the volume of the blown-up 2-cycle. Furthermore, 11-dimensional supergravity demands it to obey

\begin{equation}
d\,\beta=0\, ,\qquad d(\,h\star_8\,\beta)=0\, ;
\end{equation}
where the $\star_8$ is the Hodge-dual with respect to the 8-dimensional resolved cone metric. Following \cite{Klebanov:2007cx} it is possible to argue for the existence of such $\beta$. First, in the unwarped case $\beta$ is just a harmonic two-form. Furthermore, in the warped case the equations above can be seen to arise from an action, thus satisfying a minimum principle. 

On the other hand, the 1-form $A$ can be conveniently dualized into an scalar in the 3-dimensional field theory directions. In fact, the Hodge dual of the above 3-form potential involves

\begin{equation}
\delta\,G_7=\star_3\,dA\wedge \,h\,\star_8\,\beta\, .
\end{equation}
Defining $\star_3\,dA=d\,p$, we can write the above field strength fluctuation as

\begin{equation}
\delta\,G_7=d\,p\wedge \,h\,\star_8\,\beta\, .
\end{equation}
Thus, making use of the equations of motion above, we see that we can take $\delta\,C_6=p\,h\, \star_8\beta$. As $\beta$ is proportional, in the bottom of the cone, to the volume form of the blown-up cycle, its dual precisely goes through the $\Sigma_5$ cycle. Thus, this supergravity fluctuation couples to the baryonic condensate described above through the Wess-Zumino part of the euclidean brane action. In fact, this provides the phase of the $\mathcal{B}$ VEV, so that schematically

\begin{equation}
\langle \mathcal{B}\rangle\sim z^{-\Delta(\Sigma_5)}\, e^{i\,p}\, ;
\end{equation}
which shows that $p$ must be identified with the Goldstone boson of symmetry breaking. Indeed, we could use a different gauge for the $\delta G_7$ field strength such that assymptotically

\begin{equation}
\delta C_6\sim z\, dp\wedge {\rm Vol}(\Sigma_5)\, ;
\end{equation}
which implies $\langle J_{\mu}^B\rangle\sim \partial_{\mu}p$ for the boundary theory.

\section{An example: the cone over $Q^{111}$}\label{example}

We have been so far kept the discussion generic. Let us put the previous machinery at work in a particularly interesting example: the cone over $Q^{111}$. This is a toric $CY_4$ manifold, whose toric diagram we anticipated in (\ref{fig:toricdiagramQ111}). Its isometry group is $SU(2)^3\times U(1)_R$, and in local coordinates the explicit metric is

\begin{equation}
\label{Qiiimetric}
d s^2(Q^{111}) = \frac{1}{16}\left(d\psi+\sum_{i=1}^3 \cos\theta_id\phi_i\right)^2 + \frac{1}{8}\sum_{i=1}^3 
\left(d \theta_i^2 + \sin^2\theta_id\phi_i^2\right)~.
\end{equation}
Here $(\theta_i,\phi_i)$ are standard coordinates on three copies of $S^2=\mathbb{CP}^1$, $i=1,2,3$, and $\psi$ has period $4\pi$. The two Killing spinors are charged under $\partial_\psi$, which is dual to the $U(1)_R$ symmetry. The metric (\ref{Qiiimetric}) shows very explicitly the regular structure of a $U(1)$ bundle over the standard K\"ahler-Einstein metric on $\mathbb{CP}^1\times\mathbb{CP}^1\times\mathbb{CP}^1$, where $\psi$ is the fibre coordinate and the Chern numbers are $(1,1,1)$. 

We now consider a stack of $N$ M2 brane at the tip of this cone. The near horizon geometry is the standard Freund-Rubin type $AdS_4\times Q^{111}$. Since $b_2(Q^{111})=2$, according to the general discussion above, we should expect two vector fields in $AdS_4$ arising from KK reduction on the dual 5-cycles of $C_6$ fluctuations.

\subsection{Two versions for the same theory}

From the toric diagram in fig.(\ref{fig:toricdiagramQ111}) we can immediately read the minimal gauged linear $\sigma$-model (GLSM) realizing the variety. It contains 6 fields whose charges under the $U(1)_I\times U(1)_{II}$ gauge symmetries are

\begin{eqnarray}
\label{toricdesc}
\begin{array}{c|cccccc} & a_1 & a_2 & b_1 & b_2 & c_1 & c_2 \\ \hline U(1)_I & -1 & -1 & 1 & 1 & 0 & 0 \\ U(1)_{II} & -1 & -1 & 0 & 0 & 1 & 1 \end{array}
\end{eqnarray}

\begin{figure}[ht]
\begin{center}
\includegraphics[scale=1,angle=0]{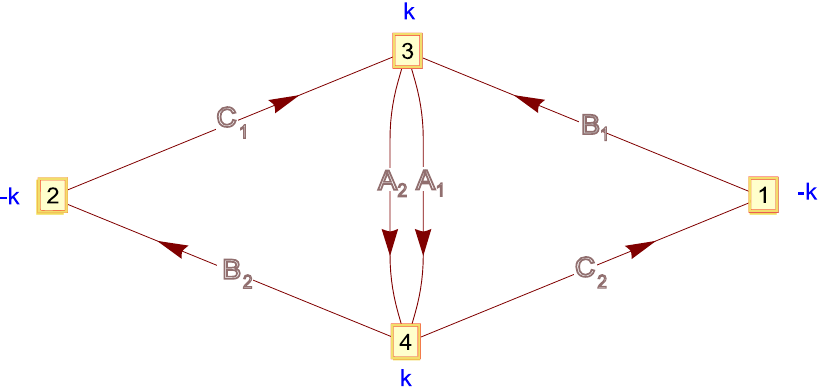}
\end{center}
\caption{The toric diagram for $\mathcal{C}(Q^{111})$.}
\label{fig:quiverQ111}
\end{figure}

Following the ABJM example, we look for a Chern-Simons matter theory where to embed this minimal GLSM. As shown in \cite{Franco:2008um}, we can succinctly encode such theory in the quiver shown in fig.(\ref{fig:quiverQ111})

We assume all the nodes to come with an $\mathcal{N}=2$ $U(N)$ Chern-Simons action with the level indicated in fig.(\ref{fig:quiverQ111}). Furthermore, the superpotential reads

\begin{equation}
W={\rm Tr}\Big(\, C_2\, B_1\, A_i\, B_2\,C_1\,A_j\,\epsilon^{ij}\,\Big)\, .
\end{equation}

It can be shown \cite{Franco:2008um} that this theory indeed contains, at $k=1$, the desired GLSM, where $a_i\leftrightarrow A_i,\,b_i\leftrightarrow B_i,\,c_i\leftrightarrow C_i$. Let us give a flavor on the proof by describing the generic construction associated to $\mathcal{N}=2$ toric Chern-Simons-matter quiver theories (see  \cite{Jafferis:2008qz, Martelli:2008si, Hanany:2008cd} for more details). For a start, we note that $\mathcal{N}=2$ SUSY in 3-dimensions can be though as the dimensional reduction along, say, $x^3$ of 4-dimensional $\mathcal{N}=1$. In particular, upon gauge fixing, the 3-dimensional vector supermultiplet contains two scalars $D,\, \sigma$ arising respectively from the 4-dimensional $D$ scalar and $A_3$ component of the gauge field. Crucially, it turns out that both scalars are auxiliary fields for Chern-Simons matter theories (see \textit{e.g.} \cite{Gaiotto:2007qi}) and thus must be integrated out. The resulting $F$ and (generalized) $D$ flatness conditions turn out to be

\begin{equation}
\label{VMSeqns}
\begin{array}{c c c c c}
 \partial_{X_{ab}} W = 0\, , & & -\sum\limits_{b=1}^G {X_{ba}}^{\dagger} {X_{ba}} + 
\sum\limits_{c=1}^G  {X_{ac}} {X_{ac}}^{\dagger} =\frac{k_a\sigma_a}{2\pi}\, , & &  \sigma_a X_{ab} - X_{ab} \sigma_b = 0\,  ;
\end{array}
\end{equation}
where latin indices run to the $G$ gauge groups (in the case at hand four) and $X_{a b}$ is a $(U(N)_a,\,U(N)_b)$ bifundamental. 

The last equation in (\ref{VMSeqns}) is automatically satisfied upon diagonalizing our fields and taking $\sigma_a=\sigma\,I_{N}\,\forall a$. Thus, the theory breaks into $N$ copies of the $U(1)$ version. Furthermore, assuming $\sum k_a=0$ it is easy to see that the equations setting $\mu_a=0$ reduce to $G-2$ independent equations. On the other hand, it is a standard result that for toric $W$ the set of $F$-flat configurations --the so-called \emph{master space}, see \textit{e.g.} \cite{Forcella:2008bb}-- is of dimension $G+2$. Thus, out of this $G+2$ dimensional master space and after imposing the $G-2$ generalized $D$ terms, we finally have a 4-dimensional toric manifold as moduli space. One can verify that for the case at hand, at $k=1$, this manifold is indeed the cone over $Q^{111}$. Let us stress that this computation merely focus on the abelian moduli space. In fact, at the abelian level the $W$ vanishes. A more detailed analysis requires the study of the chiral ring at the non-abelian level, which on general grounds must match the coordinate ring of the variety. Generically, this is a very difficult task, as we \textit{a priori} expect crucial non-perturbative effects associated to monopole operators. In order to simplify the problem, we can consider the large $k$ limit, as the dimension of such monopole operators should scale with $k$ thus decoupling. In that limit, the chiral ring is composed out of standard gauge invariant operators, \textit{i.e.} closed loops in the quiver modulo $F$-terms. Conversely, the $k\ne 1$ moduli space is indeed an orbifold of the $k=1$ variety. As shown in \cite{Franco:2009sp}, it is possible to exactly match the coordinate ring of this orbifolded variety to the non-abelian chiral ring of the theory above, in particular explicitly checking the $W$ structure. We refer to \cite{Franco:2009sp} for a complete discussion.

Let us note that the orbifold action breaks the original $SU(2)^3$ down to the single $SU(2)$ present in the superpotential. This action in fact has fixed points away from the tip of the cone. This motivated \cite{Jafferis:2009th, Benini:2009qs} to propose alternative theories containing fundamental matter associated to the flavor branes, from a IIA perspective, to which these singularities lead. We refer to these works, as well as to \cite{Cremonesi:2010ae}, for further details.

Being the gauge group of the theory we have just discussed $U(N)^4$, it cannot accommodate for gauge invariant baryon-like operators. It must then correspond to a choice of boundary conditions in the gravity dual where the 2 vector fields in $AdS_4$ arising from KK reduction on the $b_2(Q^{111})=2$ 2-cycles have $j_{\mu}=0$; that is, they are dual to boundary gauge symmetries. As discussed above, the field theory dual to changing these boundary conditions can be found by acting with the $\{\mathcal{T},\,\mathcal{S}\}$ $SL(2,\,\mathbb{Z})$ generators, as these correspond to swapping boundary conditions. In order to further proceed, let us strip off the abelian part of the gauge symmetry and call the denote the corresponding generators $\mathcal{A}_i$. We define

\begin{equation}
\mathcal{B}_k=\mathcal{A}_1+\mathcal{A}_2-\mathcal{A}_3-\mathcal{A}_4\, , \qquad \mathcal{B}_d=\mathcal{A}_1+\mathcal{A}_2+\mathcal{A}_3+\mathcal{A}_4\, ,\qquad \mathcal{A}_+=\mathcal{A}_1-\mathcal{A}_2\, ,\qquad \mathcal{A}_-=\mathcal{A}_3-\mathcal{A}_4\, .
\end{equation}
It is not hard to show that the full action, at $k=1$ can be written as (we focus on the bosonic content)

\begin{equation}
S=\frac{1}{4\pi}\, \int \mathcal{A}_+\, \wedge d\mathcal{A}_+-\frac{1}{4\pi}\, \int \mathcal{A}_-\, \wedge d\mathcal{A}_-+S_{SU}\, ,\qquad S_{SU}=\frac{1}{4\pi}\, \int \mathcal{B}_k\, \wedge d\mathcal{B}_d+S_R\, ;
\end{equation}
where  $S_{R}$ collects the remaining terms from the original lagrangian, and in particular contains $\mathcal{A}_{\pm}$ through the covariant derivatives of the fields. In fact, let us consider the theory defined by this action \textit{per se}. We note that this is an $SU(N)^4\times U(1)_k\times U(1)_d$ theory, where the abelian factors are given by the $\mathcal{B}_k,\,\mathcal{B}_d$ fields above. 

Starting from $S_{SU}$ alone, we can think of the $\mathcal{A}_{\pm}$ as background non-dynamical gauge fields. Thus, we are in the situation described in \cite{Witten:2003ya}, where we can act with the generators $\{\mathcal{S},\,\mathcal{T}\}$.\footnote{We will follow a slightly different path as in \cite{Benishti:2010jn}. We thank C.Closset and S.Cremonesi for discussions on this topic.} Let us now act with the $\mathcal{S}$ generator by adding new background gauge fields $\mathcal{C}_{\pm}$. 

\begin{equation}
S_{SU}[\mathcal{A}_+,\,\mathcal{A}_-]\rightarrow S_{SU}[\mathcal{A}_+,\,\mathcal{A}_-]+\frac{1}{2\pi}\int \mathcal{C}_+\wedge d\mathcal{A}_++\frac{1}{2\pi}\int \mathcal{C}_-\wedge d\mathcal{A}_-\, .
\end{equation}
While we won't write it explicitly, the $\mathcal{S}$ operation also introduced a functional integral over $\mathcal{A}_{\pm}$. We can act again with the $\mathcal{S}$ generator on the newly generated background gauge symmetries $\mathcal{C}_{\pm}$, so that we find, grouping terms

\begin{equation}
S_{SU}[\mathcal{A}_+,\,\mathcal{A}_-]+\frac{1}{2\pi}\int \mathcal{C}_+\wedge d(\mathcal{A}_++\mathcal{D}_+)+\frac{1}{2\pi}\int \mathcal{C}_-\wedge d(\mathcal{A}_-+\mathcal{D}_-)\, .
\end{equation}
Again, we stress that a functional integration, this time over $\mathcal{C}_{\pm}$ has been introduced. Acting now with the $\mathcal{T}$ generator on the new background gauge symmetries $\mathcal{D}_{\pm}$ we find 

\begin{equation}
S_{SU}[\mathcal{A}_+,\,\mathcal{A}_-]+\frac{1}{2\pi}\int \mathcal{C}_+\wedge d(\mathcal{A}_++\mathcal{D}_+)+\frac{1}{2\pi}\int \mathcal{C}_-\wedge d(\mathcal{A}_-+\mathcal{D}_-)+\frac{1}{4\pi}\int \mathcal{D}_+\wedge d\mathcal{D}_+-\frac{1}{4\pi}\int \mathcal{D}_-\wedge d\mathcal{D}_-\, .
\end{equation}
The functional integration over $\mathcal{C}_{\pm}$ leads to a functional $\delta$ setting $\mathcal{D}_{\pm}=-\mathcal{A}_{\pm}$, thus recovering exactly $S_U$. Thus, from this perspective, we can consider the theory defined by $S_{SU}$ as the the dual to the background with boundary conditions fixing $a_{\mu}$ in the boundary. In turn, these boundary conditions allow for wrapped M5 branes and must be dual to a theory with global baryonic symmetries. Conversely, upon considering the $S_{SU}$ theory, we no longer need to demand gauge invariance with respect to the $\mathcal{A}_{\pm}$ gauge symmetries. Thus, operators such as \textit{e.g.} ${\rm det}A_i$ become gauge-invariant and are the natural candidates for duals to the wrapped M5 branes.

We can understand the previous procedure in yet a different manner. The M5 branes corresponding to baryonic operators are in one-to-one correspondence with the divisors, encoded in the toric diagram arising from the GLSM charge matrix (\ref{toricdesc}). Thus, that particular combination of $U(1)$'s naturally encodes the baryonic charges necessary to describe all baryonic operators. In turn, the Chern-Simon-matter theory described above contains precisely this GLSM. In fact, the sequence of $\{\mathcal{T},\,\mathcal{S}\}$ operations above amount to ungauge precisely these two $U(1)$'s (which are nothing but $\mathcal{A}_+\pm\mathcal{A}_-$).

\subsection{Spontaneous symmetry breaking}

As discussed, spontaneous symmetry breaking amounts to resolution in the gravity dual. In \cite{Benishti:2010jn} a comprehensive algebraic analysis of the cone over $Q^{111}$ was performed, paying attention in particular to the space of Kahler parameters which account for the resolutions. From the point of view of the GLSM above, by turning on Fayet-Ilopoulos parameters we can achieve every possible resolution of the geometry. In turn, for each of the resolutions of $\mathcal{C}(Q^{111})$, there is a corresponding Ricci-flat K\"ahler metric that is asymptotic to the cone metric over $Q^{111}$. More precisely, there is a unique such metric for each choice of K\"ahler class, or equivalently FI parameter $\zeta_1,\zeta_2\in \mathbb{R}$. Roughly speaking, these parameteres correspond to the volumes of the 2 2-cycles which can be blown up. Denoting the radii of these blown-up $S^2$'s by $(a,\,b)$, the resolved Calabi-Yau metric is  given by
\begin{eqnarray}\label{resolvedQ111}
d s^2(X) &=&\kappa(r)^{-1} d r^2+\kappa(r)\frac{r^2}{16}\Big(d\psi+\sum_{i=1}^3 \cos\theta_i d\phi_i\Big)^2+\frac{(2a+r^2)}{8}\Big(d\theta_2^2+\sin^2\theta_2 d\phi_2^2\Big)\nonumber \\ &&+\frac{(2b+r^2)}{8}\Big( d\theta_3^2+\sin^2\theta_3 d\phi_3^2\Big)+\frac{r^2}{8}\Big(d\theta_1^2+\sin^2\theta_1 d\phi_1^2\Big)~,
\end{eqnarray}
where 
\begin{equation}\label{kappa}
\kappa(r)=\frac{(2A_-+r^2)(2A_++r^2)}{(2a+r^2)(2b+r^2)}~,
\end{equation}
$a$ and $b$ are arbitrary constants determining the sizes of the blown-up $S^2$'s; and we have also defined 
\begin{equation}
A_\pm=\frac{1}{3}\Big(2a+2b\pm\sqrt{4a^2-10ab+4b^2}\Big)~.
\end{equation}
We are interested in studying supergravity backgrounds corresponding to M2 branes localized on one of these resolutions of $\mathcal{C}(Q^{111})$. If we place $N$ spacetime-filling M2 branes at a point $y\in X$, we must then solve the following equation for the warp factor

\begin{eqnarray}
\Delta_x h[y] = \frac{(2\pi \ell_p)^6N}{\sqrt{\det g_X}} \delta^8(x-y) \ ,
\end{eqnarray}
where $\Delta$ is the scalar Laplacian on the resolved cone. In order to simplify the problem, let us analyse the case in which we partially resolve the cone,  setting $a=0$ and $b>0$. With no loss of generality, we put the $N$ M2 branes at the north pole of the blown-up $S^2$ parametrized by $(\theta_3,\phi_3)$. We then find

\begin{eqnarray}
h(r,\theta_3) &=&\sum_{l=0}^\infty\,H_{l}(r)\, P_{l}(\cos\theta_3)~,\nonumber \\
H_{l}(r)&=& \mathcal{C}_{l}\, \Big(\frac{8b}{3r^2}\Big)^{3(1+\beta)/2}\, _2F_1\left(-\frac{1}{2}+\frac{3}{2}\beta,\frac{3}{2}+\frac{3}{2}\beta,1+3\beta,-\frac{8b}{3r^2}\right)~,
\label{Q111-warp-factor}
\end{eqnarray}
where $P_{l}$ denotes the $l$-th Legendre polynomial, 
\begin{equation}
 \beta=\beta(l)=\sqrt{1+\frac{8}{9}l(l+1)}~,
\end{equation}
and the normalization factor $\mathcal{C}_{l}$ is given by
\begin{eqnarray}
\mathcal{C}_{l}&=&\frac{3\Gamma(\frac{3}{2}+\frac{3}{2}\beta)^2}{2\Gamma(1+3\beta)}\left(\frac{3}{8b}\right)^3\, (2l+1)\, R^6~,\\\label{anothereqn}
 R^6&=&\frac{(2\pi\ell_p)^6 N}{6{\rm vol}(Q^{111})}=\frac{256}{3}\pi^2 N \ell_p^6~.
\end{eqnarray}
In the field theory this solution corresponds to breaking one combination of the two global $U(1)$ baryonic symmetries, rather than both of them. As discussed in general above, the resolution of the cone can be interpreted in terms of giving an expectation value to a certain operator $\mathcal{U}$ in the field theory. This operator is contained in the same multiplet as the current that generates the broken baryonic symmetry, and couples to the corresponding $U(1)$ gauge field in $AdS_4$. Since a conserved current has no anomalous dimension, the dimension of $\mathcal{U}$ is uncorrected in going from the classical description to supergravity \cite{Klebanov:1999tb}. According to the general $AdS/CFT$ prescription \cite{Klebanov:1999tb}, the VEV of the operator $\mathcal{U}$ is dual to the subleading correction to the warp factor. For large $r$ one can show 
\begin{equation}
h(r,\theta_3)\sim \frac{R^6}{r^6}\left(1+\frac{18b\, \cos\theta_3}{5r^2}+\cdots\right)~.
\end{equation}
In terms of the $AdS_4$ coordinate $z= r^{-2}$ we have that the leading correction is of order $z$, which indicates that the dual operator $\mathcal{U}$ is dimension 1. This is precisely the expected result, since this operator sits in the same supermultiplet as the broken baryonic current, and thus has a protected dimension of 1. Furthermore, its VEV is proportional to $b$, the metric resolution parameter, which reflects the fact that in the conical ($AdS$) limit in which $b=0$ this baryonic current is not broken, and as such $\langle \mathcal{U}\rangle=0$.

Furthermore, we can compute, following the steps described for the general case, the VEV of the baryonic condensate as the volume of an euclidean brane wrapping the cone over $\Sigma_5$. While the details of the computation can be seen in \cite{Benishti:2010jn}, here we content ourselves with quoting the result

\begin{equation}
e^{-S(r_c)}=e^{{7N}/{18}} \left(\frac{8b}{3\,r_c^2}\right)^{\frac{N}{3}}\left(\sin \frac{\theta_3}{2}\right)^{N}~;
\label{vev}
\end{equation}
where $r_c$ is the radial cut-off. From (\ref{vev}) we can read off the dimension of the associated baryonic operator $\Delta(\mathcal{B})=\frac{N}{3}$, which suggests that, if $\mathcal{B}={\rm det}X$, then $\Delta(X)=\frac{1}{3}$. In fact, in accordance with the results in \cite{Fabbri:1999hw}, a similar computation shows that all baryonic operators must have the same scaling dimension. In turn, in the context of the Chern-Simons-matter quiver gauge theory described in the previous subsection, this implies that all fields have the same $\Delta=1/3$ scaling dimension, and hence $R=\frac{1}{3}$. This is in fact consistent with the sextic superpotential, as this assignation of R-charges ensures it to be marginal at the putative fixed point. In fact, in view of these results it would be very interesting to apply the recently discovered techniques of \cite{Jafferis:2010un} along the lines of the appendix for the $Q^{111}$ theory to confirm or disprove its potential agreement. We leave this as an open question for future work.

\section{Conclusions} 

Global symmetries are important tools in studying the spectrum of a gauge theory. In the context of the $AdS/CFT$ duality a particularly important set of such symmetries are those which arise from KK reduction of the supergravity $p$-forms in non-trivial cycles yielding to $AdS$ vectors. Following the terminology of the $AdS_5$ case, we dubbed such symmetries as baryonic.  These symmetries appear as particularly interesting and important in the $AdS_4/CFT_3$ case, as they behave much differently from the $AdS_5$ case. In particular, on the gravity side, the two possible fall-offs are admissible, thus leading to two possible $AdS_4/CFT_3$ dualities depending on the chosen boundary conditions. In turn, in the field theory side, these correspond to a choice of gauged \textit{vs.} global baryonic symmetry.

As briefly mentioned, the $CY_4$'s of interest can also potentially contain 6-cycles. While they are not directly related to the baryonic symmetries we discussed --as they do not yield to vectors in $AdS_4$ upon KK reduction of $p$-forms--, it would be very interesting to clarify their role; as they might lead to non-perturbative, instantonic, corrections to the superpotentials. We refer to \cite{Benishti:2010jn} for a first study along these lines.

While a lot has been learned recently about the $AdS_4\times CFT_3$ duality, much remains yet to be clarified, specially from the field theory perspective in the $\mathcal{N}=2$ case. In particular, the gravity analysis briefly reviewed above following \cite{Benishti:2010jn} must yield to important consistency checks. As we described, in the particular $\mathcal{C}(Q^{111})$ case described, the gravity predictions are in fact consistent with the expectations for the theory proposed in \cite{Franco:2008um}. Nevertheless, it still remains to perform a conclusive $\mathcal{Z}$ minimization analysis in the spirit of that in the appendix. Very recently a series of very refined checks involving the superconformal index have been performed in \cite{Imamura:2011uj, Cheon:2011th}. While flavored theories appear better behaved, the full picture yet remains to clarified. We leave such analysis as an open problem for the future.

\section*{Acknowledgements}

The author would like to thank AHEP and the guest editors --specially Yang-Hui He-- of the for the special issue on \textit{Computational Algebraic Geometry in String and Gauge Theory} for the invitation to write this contribution. The author is supported by the Israel Science Foundation through grant 392/09. He wishes to thank N.Benishti, A.Hanany, S.Franco, I.Klebanov, J.Park and J.Sparks for many discussions and explanations, as well as for very enjoyable collaborations. He also would like to thank C.Closset and S.Cremonesi for enlightening discussions.

\begin{appendix}

\section{$\mathcal{Z}$-minimization for HVZ}

Following \cite{Jafferis:2010un}, the properties of the putative fixed point of a 3d theory are encoded in the minimization of the modulus squared of the partition function regarded as a function of the trial R-charges (which in 3d are equal, at the SCFT point, to the scaling dimensions). As the theories which we consider do not break the parity symmetry, the partition function itself is real, and thus it is enough to minimize it. Following the localization procedure in \cite{Jafferis:2010un, Hama:2010av}, one can check that for a generic quiver theory with gauge group $U(N)^G$ and a number of bifundamental fields $X$ in the $(\Box_{\alpha_X},\,\bar{\Box}_{\beta_X})$ under the $\alpha_X,\, \beta_X$ factors and with trial scaling dimension $\Delta_X$, the partition function on the $S^3$ can be written as 

\begin{equation}
\mathcal{Z}=\frac{(-1)^{N\, G}}{N!^G}\, \int \, \prod_{g=1}^G\, \prod_{\alpha_g} du^g_{\alpha_g}\, e^{i\,\pi\, k_g\, (u^g_{\alpha_g})^2}\, \prod_{\alpha_g< \beta_g}\, \sinh^2(\pi\,(u_{\alpha_g}^g-u_{\beta_g}^g))\, \prod_{X}\, \prod_{\alpha_{X},\, \beta_{X}}^N\, e^{\ell(1-\Delta_{X}+i\,(u^i_{\alpha_{X}}-u^f_{\beta_{X}}))}\, .
\end{equation}
Let us now compare the HVZ and the ABJM theories. In order to simplify the computations, let us just focus on the $U(2)\times U(2)$ case. After some algebra, the ABJM partition function reads (we refer to \cite{Jafferis:2010un, Hama:2010av} as well to the pioneering papers on 3d localization \cite{Kapustin:2009kz, Kapustin:2010xq} for the definition of the special function $\ell$)

\begin{equation}
\mathcal{Z}_{ABJM}^{U(2)}=\frac{1}{4\,k} \int\, dx\, dy\, e^{i\,2\,\pi\, k\, (x^2-y^2)}\, \sinh^2(2\,\pi\,x)\,\sinh^2(2\,\pi\,y)\, e^{f( x,\, y)}\, ,
\end{equation}
with

\begin{equation}
f(x,\,y)=2\sum_{s_1=\pm,\, s_2=\pm}\ell(\Delta+s_1\, (x+s_2\,y))+\ell(1-\Delta+s_1\, (x+s_2\,y))\, .
\end{equation}
In order to obtain these expressions we made use of the constraints imposed by the superpotential, which allows to express all dimensions as a function of a single one $\Delta$. As expected, the partition function is minimized at $\Delta=1/2$, which leads to

\begin{equation}
\label{ZABJM}
\mathcal{Z}_{ABJM}^{U(2)}=\frac{1}{2^{10}\, k}\, \int\, dx\, dy\, e^{i\,2\,\pi\, k\, (x^2-y^2)}\, \frac{\sinh^2(2\,\pi\,x)\,\sinh^2(2\,\pi\,y)}{\cosh^4(\pi\, (x+y))\, \cosh^4(\pi\, (x-y))}\, .
\end{equation}
On the other hand, for HVZ, we obtain

\begin{equation}
\mathcal{Z}_{HVZ}^{U(2)}=\frac{1}{4\,k} \int\, dx\, dy\, e^{i\,2\,\pi\, k\, (x^2-y^2)}\, \sinh^2(2\,\pi\,x)\,\sinh^2(2\,\pi\,y)\, e^{f(x,\, y)}\, ,
\end{equation}
with 

\begin{equation}
f(x,\,y)=2\sum_{s_1=\pm}\sum_{s_2=\pm}\ell(1-\Delta+i\, s_1\, (x+s_2\,y) +4\,\ell(\Delta)+2\sum_{s=\pm}\ell(\Delta+i\, 2\, s\, x)\, .
\end{equation}

While this expression is very similar to the ABJM expression, it is not quite the same. In fact, while it is minimized at $\Delta=1/2$ --leading to the R-charge assignation guessed in \cite{RodriguezGomez:2009ae}--, the final expression becomes  

\begin{equation}
\mathcal{Z}_{HVZ}^{U(2)}=\frac{1}{2^{10}\,k} \int\, dx\, dy\, e^{i\,2\,\pi\, k\, (x^2-y^2)}\, \frac{\sinh^2(2\,\pi\,x)\,\sinh^2(2\,\pi\,y)}{ \cosh^2(\pi\,(x+y))\,\cosh^2(\pi\,(x-y))\,\cosh^2(\pi\,x)}\, ,
\end{equation}
which is just different from the ABJM result (\ref{ZABJM}) for all $k$. We note however, that the same computation for $U(1)\times U(1)$ indeed gives the same answer for the two theories. 

\end{appendix}

\end{document}